\documentclass[final]{article}

\usepackage{arxiv}

\usepackage[utf8]{inputenc} 
\usepackage[T1]{fontenc}    
\usepackage{url}            
\usepackage{booktabs}       
\usepackage{amsfonts}       
\usepackage{nicefrac}       
\usepackage{microtype}      
\usepackage{lipsum}
\usepackage[range-phrase=--,range-units=single,separate-uncertainty,multi-part-units = single,detect-weight=true,]{siunitx}
\usepackage{textcomp} 

\usepackage{nomencl}
\makenomenclature

\usepackage{ifthen}
\renewcommand{\nomgroup}[1]{%
  \ifthenelse{\equal{#1}{A}}{ {\vskip 4mm} \item[\textbf{Roman Letters}]}{%
  \ifthenelse{\equal{#1}{G}}{ {\vskip 4mm} \item[\textbf{Greek Letters}]}{%
  \ifthenelse{\equal{#1}{M}}{ {\vskip 4mm} \item[\textbf{Mathematical Notations}]}{%
  \ifthenelse{\equal{#1}{S}}{ {\vskip 4mm} \item[\textbf{Subscripts and Superscripts}]}{%
  }}}}%
}

\usepackage{amsmath}
\usepackage{amssymb}
\usepackage{graphicx} 
\usepackage{dcolumn} 
\usepackage{bm} 
    
\newcommand{\ttd}[2]{%
  \frac{\mathrm{d}^{2}#1}{\mathrm{d}#2^{2}}%
  }

\newcommand{\ppd}[2]{%
  \frac{\partial^{2}#1}{\partial#2^{2}}%
  }
  
\usepackage{blindtext}
\usepackage{xcolor}
\usepackage[title]{appendix}
\usepackage{cite}
\usepackage{stfloats} 
\usepackage{hyperref}
\usepackage{cleveref}

\crefname{equation}{Eq.}{Eqs.}
\Crefname{equation}{Equation}{Equations}

\crefname{table}{Table}{Tables}
\Crefname{table}{Table}{Tables}

\crefname{figure}{Fig.}{Figs.}
\Crefname{figure}{Figure}{Figures}

\crefname{section}{Sec.}{Secs.}
\Crefname{section}{Section}{Sections}

\crefname{chapter}{Chap.}{Chaps.}
\Crefname{section}{Section}{Sections}

\title{Heterogeneous Angular Spectrum Method for Trans-skull Imaging and Focusing}

\author{
  Scott Schoen Jr\\\
  Woodruff School of Mechanical Engineering\\
  Georgia Institute of Technology \\
  Atlanta, GA USA \\
  \texttt{scottschoenjr@gatech.edu} \\
   \And
 Costas D. Arvanitis \\
  Woodruff School of Mechanical Engineering and Coulter Department of Biomedical Engineering\\
  Georgia Institute of Technology and Emory University\\
  Atlanta, GA, USA \\
  \texttt{costas.arvanitis@gatech.edu} 
}

\begin{document}
\maketitle
\frenchspacing

\vspace{-5mm}
\begin{center}
  \begin{minipage}[]{0.685\textwidth}
    \small\textbf{%
    © 2020 IEEE.
    Personal use of this material is permitted.
    Permission from IEEE must be obtained for all other uses, in any current or future media, including reprinting/republishing this material for advertising or promotional purposes, creating new collective works, for resale or redistribution to servers or lists, or reuse of any copyrighted component of this work in other works.
    Article DOI: 1010.1109/TMI.2019.2953872.
    }
  \end{minipage}
\end{center}
\vspace{5mm}

\begin{abstract}
Ultrasound, alone or in concert with circulating microbubble contrast agents, has emerged as a promising modality for therapy and imaging of  brain diseases. 
While this has become possible due to advancements in  aberration correction methods, a range of applications, including adaptive focusing and tracking of the microbubble dynamics through the human skull, may benefit from even more computationally efficient methods to account for skull aberrations. 
Here, we derive a general method for the angular spectrum approach (ASA) in a heterogeneous medium, based on a numerical marching scheme to approximate the full implicit solution.
We then demonstrate its functionality with simulations for (human) skull-related aberration correction and trans-skull passive acoustic mapping. 
Our simulations show that the general solution provides accurate trans-skull focusing as compared to the uncorrected case (error in focal point location of \SI{1.0 \pm 0.4}{\milli\meter} vs \SI{2.2 \pm 0.7}{\milli\meter}) for clinically relevant frequencies (\SIrange{0.25}{1.5}{\mega\hertz}), apertures (\SIrange{50}{100}{\milli\meter}), and targets, with peak focal pressures approximately \SI{30 \pm 17}{\percent} of the free field case, with the effects of skull attenuation and amplitude shading included.  
In the case of source localization, our method leads to an average of \SI{75}{\percent} error reduction (from \SI{2.9 \pm 1.8}{\milli\meter} to \SI{0.7 \pm 0.5}{\milli\meter}) and \SIrange{40}{60}{\percent} increase in peak intensity, evaluated over the range of frequencies (\SIrange{0.4}{1.2}{\mega\hertz}), apertures (\SIrange{50}{100}{\milli\meter}), and point source locations (40~mm by 50~mm grid) as compared to the homogeneous medium ASA.
Overall, total computation times for both focusing and point source localization of the order milliseconds (\SI{166 \pm 37 }{\milli\second}, compared with \SI{44 \pm 4}{\milli\second} for the homogeneous ASA formulation) can be attained with this approach. 
Collectively our findings indicate that the proposed phase correction method based on the ASA could provide a computationally efficient and accurate method for trans-skull transmit focusing and imaging of point scatterers, potentially opening  new possibilities for treatment and diagnosis of brain diseases.
\end{abstract}


\section{Introduction}
\label{sec:Introduction}
Ultrasound has emerged as a novel modality for the treatment and imaging of brain diseases and disorders \cite{demene_functional_2017,elias_randomized_2016}. 
When enhanced by circulating microbubble contrast agents---lipid, albumin, or polymer-shelled gas pockets (\SIrange{1}{10}{\micro\meter}) that scatter sound and vibrate in response to incident ultrasound---it can enable a range of new therapeutic interventions\cite{auboire_microbubbles_2018,arvanitis_mechanisms_2018,idbaih_safety_2019,arvanitis_cavitation-enhanced_2016,arvanitis_bloodbrain_2019,jones_advances_2019} and open new possibilities for imaging \cite{demene_transcranial_2019,errico_ultrafast_2015,oreilly_super-resolution_2013}.
This has been facilitated by advancements in  methods to account for skull aberrations. 

In their simplest form, aberrations introduced by the skull may be corrected with knowledge of phase and amplitude at the transmitter due to a point source at the target, that can then be conjugated to achieve trans-skull focusing on transmit.
Such phase and amplitude modulations may be determined experimentally via measurement of the field due to an induced cavitation source at the target location inside the skull \cite{pernot_ultrasonic_2006,haworth_towards_2008,gateau_transcranial_2010}, however cavitation events in the brain are not always desirable, and the relatively high cavitation threshold requires additional corrections or high power arrays \cite{kyriakou_review_2014}.
Use of corrections based on analysis of time domain signals measured from contrast agent microbubble emissions has been demonstrated\cite{oreilly_investigating_2014}, though the location of these cavitation events cannot be controlled precisely.
Alternatively, a point source can be placed within the skull to enable effective trans-skull therapy
\cite{thomas_ultrasonic_1996,hynynen_demonstration_1998,white_transcranial_2005,pernot_vivo_2007}; however this technique requires invasive placement at the target, which is not possible in realistic scenarios.
For these reasons, several simulation techniques have been proposed for aberration correction, including finite difference time domain (FDTD) \cite{aubry_experimental_2003,pulkkinen_numerical_2014} and k-space propagation models\cite{clement_non-invasive_2002,tabei_simulation_2003,mcdannold_transcranial_2010}. 
In these methods, the simulation of the acoustic propagation due to a point source inside the brain is used to estimate the required phases for aberration correction. 
While these methods can be very precise, as they may account for a broad range of physical effects including, e.g., mode conversion, viscosity and nonlinearity, there exists a fundamental trade-off between their accuracy and required computational complexity \cite{jones_comparison_2015}. 

Currently an effective trade-off is obtained by the use of a stepwise, locally homogeneous angular spectrum approach (ASA), which employs either coordinate rotations to account for refraction\cite{clement_non-invasive_2002} or accounts for spatial heterogeneity in the space domain before spectral propagation\cite{vyas_ultrasound_2012}.
While these ASA implementations currently provide a reasonable compromise between computational complexity and aberration correction, they do not solve intrinsically the full wave propagation problem \cite{vyas_ultrasound_2012,vyas_transcranial_2014,leung_rapid_2019,jones_comparison_2015}. 

Inherently related to the problem of errors in trans-skull FUS targeting due to skull aberrations is that of source localization.
This problem has emerged as a priority in microbubble enhanced US therapy and imaging: when exposed to ultrasound, microbubbles scatter diverging pressure waves that, due to their size (several orders of magnitude smaller than the wavelength), act as point sources. 
The information carried by these waves can enable the spatiotemporal characterization of the microbubble dynamics. 
During therapeutic interventions, this information is used to ensure that the desired type of oscillation (i.e., stable or inertial) is taking place at the intended location (i.e. treatment monitoring) \cite{arvanitis_passive_2017}. 

Several approaches based on passive acoustic mapping (PAM) techniques have been proposed for monitoring the microbubble dynamics. 
While direct implementations do not account for aberrations \cite{gyongy_passive_2010,coviello_passive_2015,haworth_quantitative_2017,arvanitis_passive_2017}, recent work has successfully incorporated human skull aberration corrections into both frequency-domain\cite{jones_transcranial_2013} and time-domain \cite{oreilly_super-resolution_2013,oreilly_three-dimensional_2014,jones_comparison_2015,deng_multi-frequency_2016,arvanitis_transcranial_2015,jones_three-dimensional_2018} PAM implementations. 
These methods require either measurement (invasive) \cite{thomas_ultrasonic_1996,gateau_transcranial_2010}, simulations to calculate delays for each point in the image, or a representative point to be used for a local region \cite{jones_experimental_2015,jones_comparison_2015}, which incurs a computational expense that scales with the field of view.
These methods are thus either infeasible or time consuming (up to several minutes) unless GPUs are used to speed up the computations \cite{jones_comparison_2015}.
This problem can potentially be overcome with the ASA method\cite{arvanitis_passive_2017}, however current implementations do not account for aberrations.

The spectral selectivity inherent to frequency domain methods such as the ASA \cite{haworth_quantitative_2017,arvanitis_passive_2017} is important for characterizing the type of oscillation (e.g. harmonics vs broadband). 
However current implementations assume a homogeneous medium. 
In addition to spectral selectivity, reconstructions on the order of milliseconds are important for closed-loop control of the  microbubble dynamics\cite{patel_closed_2018} and improved temporal resolution during microbubble imaging.\cite{oreilly_three-dimensional_2014,errico_ultrafast_2015} 
While time domain methods \cite{gyongy_passive_2010,coviello_passive_2015,jones_three-dimensional_2018} can readily incorporate corrections and extract for certain frequency content via filtering, they generally incur high computational loads, unless GPU units are used to speed up the computations \cite{jones_three-dimensional_2018}.
Hence, fast and frequency selective reconstructions to visualize the cerebrovascular microbubble dynamics through the skull may have important implications for image guided therapy and imaging \cite{jones_advances_2019}.

Herein we derive the general solution for the heterogeneous ASA. 
First, we present the derivation of a fast phase correction method for arbitrarily distributed, weakly heterogeneous medium. 
Then, through simulated acoustic propagation, we show the numerical implementation of the algorithm for focal aberration correction and frequency selective passive mapping of point sources through a human skull.
Next, we evaluate the focal aberration error, focal pressure, and spot size, as well as the point source localization error and intensity, compared with the corresponding uncorrected cases. Finally, we  evaluate the computational cost of the different algorithm permutations.

\section{Theory}

\subsection{Angular Spectrum Approach}
\label{sec:AngularSpectrum}
The angular spectrum $P$ of a monochromatic\footnote{%
A time convention of $p = \tilde{p}\,e^{-i\omega t}$ is used, so that the forward temporal transform uses a kernel of $e^{+i\omega t}$, and that of the forward spatial transform is then $e^{-i(k_{x}x + k_{y}y)}$.
The factors of $1/2\pi$ are applied on the inverse transforms.%
} pressure field $\tilde{p}$ with angular frequency $\omega$ is given by its 2D spatial Fourier transform 
\begin{align}
  P(k_{x}, k_{y}, z) 
  &= 
  \mathcal{F}_{k}[\,\tilde{p}(x, y, z)\,] 
  \nonumber \\
  &\equiv 
  \iint_{-\infty}^{\infty}{ \tilde{p}(x, y, z)e^{-i\left(k_{x}x + k_{y}y \right)}\,\mathrm{d}x\,\mathrm{d}y}\,.
  \label{eqn:AngularSpectrumDefinition}
\end{align}
Applying the spatial transform to the homogeneous Helmholtz equation $(\nabla^{2} + k^{2})\tilde{p} = 0$ yields an ordinary differential equation for the angular spectrum
\begin{align}
  \ttd{P}{z} + k_{z}^{2}P = 0\,,
  \label{eqn:TransformedAsOde}
\end{align}
where $k_{z}^{2} = (\omega/c_{0})^{2} - k_{x}^{2} - k_{y}^{2}$, and $c_{0}$ is the (constant) small signal sound speed. 
If $P_{0}$ is known at some reference plane $z = 0$, and if there are no backward-travelling waves, then \cref{eqn:TransformedAsOde} has the solution
\begin{align}
  P &= P_{0}e^{ik_{z}z}\,.
  \label{eqn:AngularSpectrumTransferFunction}
\end{align}
The acoustic field in any plane may then be reconstructed with \cref{eqn:AngularSpectrumTransferFunction} and evaluation of the inverse transform.

\subsection{Extension to Heterogeneous Media}
\label{sec:NonuniformMedia}
In the case of relatively weak heterogeneity---i.e. the sound speed $c(\boldsymbol{r})$ changes slowly compared with the wavelength---the governing equation is
\begin{align}
  \ttd{P}{z} + k_{z}^{2}P 
  &= 
  \Lambda * P\,.
  \label{eqn:InhomogeneousGoverningEquation}
\end{align}
Here, $\Lambda = \mathcal{F}_{k}\left[ k_{0}^{2}(1 - \mu )\right]$, $k_{0} = \omega/c_{0}$, $\mu = c_{0}^{2}/c^{2}$, $c_{0}$ is a reference (average) sound speed, and $*$ denotes a 2D convolution over the wavenumber components $k_{x}$ and $k_{y}$.
From \cref{eqn:TransformedAsOde}, the heterogeneity appears as a source term in the governing equation. 
In the general case, the implicit solution of \cref{eqn:InhomogeneousGoverningEquation} may be obtained with a Green's function technique
\begin{align}
  P 
  &=
  P_{0}e^{ik_{z}z}
  +
  \frac{\,e^{ik_{z}z}}{2ik_{z}}\int_{0}^{z}{ e^{-ik_{z}z'}\left( \Lambda * P  \right)\,\mathrm{d}z' }\,.
  \label{eqn:AngularSpectrumSolutionImplicit}
\end{align}
A numerical approximation of \cref{eqn:AngularSpectrumSolutionImplicit} may be made to compute $P$ at arbitrary $z$ via
\begin{align}
  P^{n + 1} 
  \approx 
  P^{n}e^{ik_{z}\Delta z}
  +
  \frac{e^{ik_{z}\Delta z}}{2ik_{z}}\,\left( P^{n} * \Lambda  \right) \times \Delta z\,,
  \label{eqn:AngularSpectrumMarchingAlgorithm}
\end{align}
where $P^{n} = P(k_{x}, k_{y}, n\Delta z)$.
See Appendix~\ref{sec:Derivations} for derivations of \cref{eqn:InhomogeneousGoverningEquation,eqn:AngularSpectrumSolutionImplicit,eqn:AngularSpectrumMarchingAlgorithm}.
Previously, Gu and Jing persented a general forward propagation scheme for field simulations that includes nonlinearity and attenuation \cite{gu_numerical_2018}.
In the absence of these effects, Eq.~(12) of that reference becomes (with the notation of this paper)
\begin{align}
  M &= \mathcal{F}_{k}\left\{%
    \left[ k_{0}^{2}\left( 1 - \frac{c_{0}^{2}}{c^{2}} \right) \right] \times \tilde{p}
    \right\}
  = \Lambda * P\,.
\end{align}
Substitution of this value of $M$ into their Eq.~(11) recovers \cref{eqn:AngularSpectrumSolutionImplicit} obtained here, indicating the consistency of the results. 

\section{Methods}

\subsection{Simulations}
\label{sec:Simulations}
Acoustic propagation (either to evaluate the transmit focusing achieved with the computed time delays or to acquire the RF data to be used with the corrected beamforming) was simulated in k-Wave \cite{treeby_k-wave:_2010}. 
As the incident angles were largely below the critical angle for skull bone ($\sim\SI{26}{\degree}$\cite{clement_enhanced_2004}), elastic effects were not included in the simulation.
The skull and tissue densities were computed from CT data of a human skull, and the spatially-dependent sound speed and absorption coefficient $\alpha$ were computed from the Hounsfield scale conversion as described previously  \cite{aubry_experimental_2003,arvanitis_transcranial_2015}.
Absorption was included via a power law model $\alpha(\boldsymbol{r}) = \alpha_{0}(\boldsymbol{r})\cdot f^{\beta}$ \cite{treeby_modeling_2010}.
A power law exponent of $\beta = 1.2$  was defined for the entire medium \cite{cobbold_foundations_2007}, as k-Wave's implementation does not allow spatial variation of this parameter.
Spatial grid spacing \SI{200}{\micro\meter} and time step \SI{40}{\nano\second} were used for all simulations (CFL number $c\Delta t/\Delta x = 0.44$), and 2000 time steps (\SI{80}{\micro\second}) were simulated.
The reference sound speed was taken to be the average value over the entire grid.
All reconstruction and aberration correction scripts were written in MATLAB and run on a standard desktop computer (Intel Core i7, four cores at 2.8~GHz and 16~GB memory) without parallel or graphical processing techniques.
The general computational flows for focal aberration correction and source localization are shown in \cref{fig:FocusingImplementation} and \cref{fig:PAMImplementation}, respectively.
While the correction can function in 3 dimensions [see, e.g., \cref{fig:PAMImplementation}(c)], due to the large number of simulations required, simulations were performed in 2D (\SI{60}{mm} by \SI{80}{mm}, with 10 point perfectly matched layer) for computational efficiency. Thus in the reconstructions $k_{y} = 0$.

\subsection{Numerical Implementation}
\label{sec:Implementation}

\subsubsection{Transmit Focusing}
\begin{figure}[!b]
    \centering
    \includegraphics[width=0.45\textwidth]{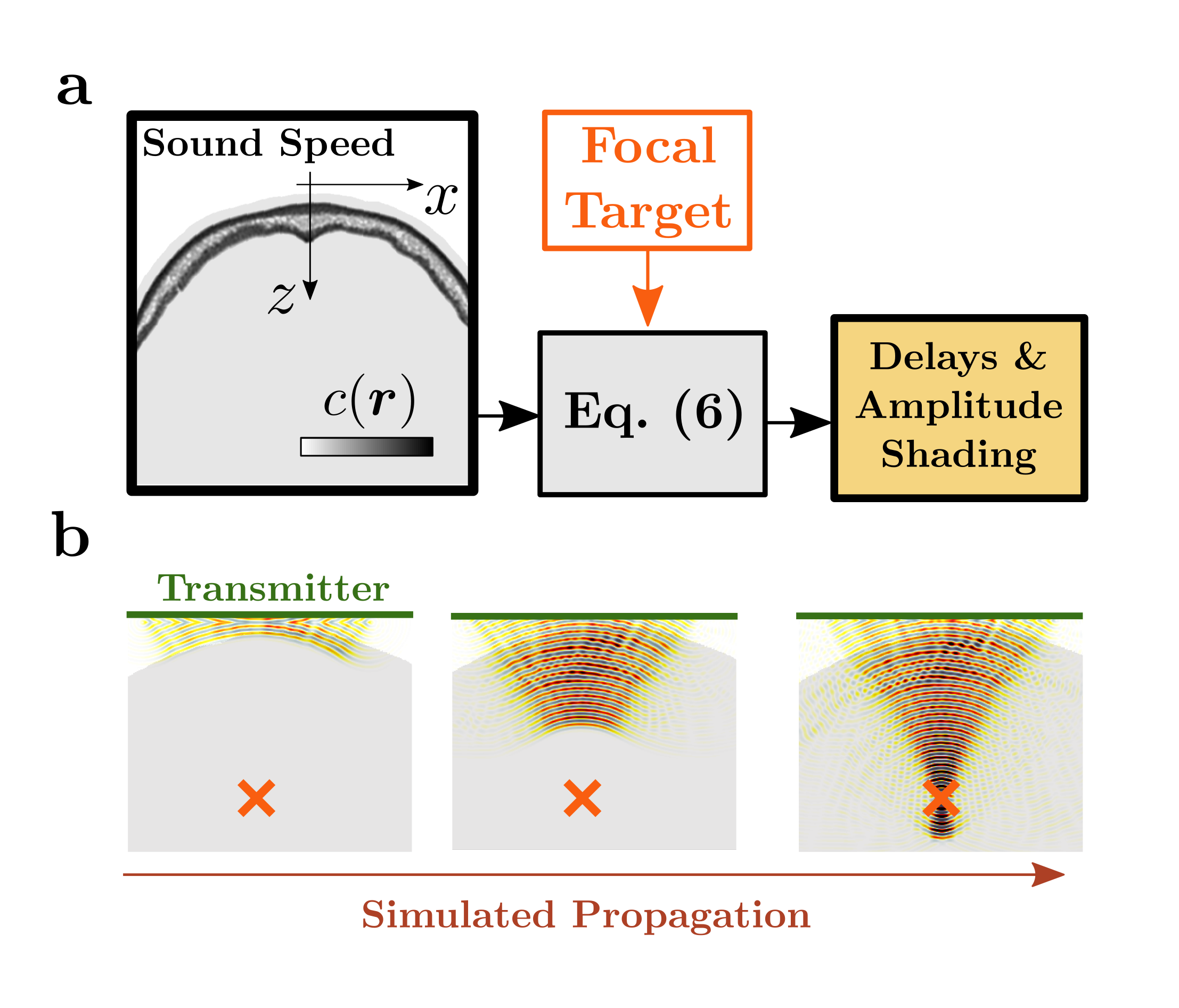}
    \caption{%
      Flow of computations for focal aberration correction.
      \textbf{(a)}~The known sound speed field $c(\boldsymbol{r})$ is used, together with the target focal spot, to compute the element amplitudes and time delays.
      \textbf{(b)}~The delays and amplitudes are then used to simulate propagation and the location of peak intensity is compared with the target point.
    }
    \label{fig:FocusingImplementation}
\end{figure}
To determine the improvement in focal aberration correction through human skull with the proposed general solution (heterogeneous ASA), the focusing delays were computed with and without phase corrections for clinically relevant frequencies (\SIrange{0.25}{1.5}{\mega\hertz}) \cite{elias_randomized_2016, idbaih_safety_2019, marquet_non-invasive_2013,wu_efficient_2018}.
To achieve focusing at a desired depth $d$, $P(k_{x}, d)$ was computed for a delta function at the desired focus, i.e., for $P_{0} = \mathcal{F}_{k}\left[ \delta( x - x_{0}, 0 ) \right]$.
To avoid very steep changes on the spatial computation grid for the corrected case, this delta function was approximated as a cosine-windowed Gaussian distribution with full width at half maximum of \SI{1.5}{mm}.
Since the field is monochromatic, the excess phase represents the total time delay $\tau$ for an array element as a function of its position along the transducer face
\begin{align}
    \tau(x) = \frac{\operatorname{arg}{\tilde{p}(x,  d)}}{\omega}\,.
    \label{eqn:FocusingDelays}
\end{align}
Phase unwrapping was used to preserve the full phase of $P$ at the transducer location.
For reference, the results of focusing using the delays calculated with \cref{eqn:FocusingDelays} were compared with the results obtained without accounting for medium heterogeneity (i.e., geometrical focusing).
In this case, the delays were computed with $c_{0} = \SI{1500}{m/s}$ directly from the target point via
\begin{align}
    \tau(x) = \frac{\sqrt{(x - x_{0})^{2} + d^{2}}}{c_{0}}\,.
    \label{eqn:GeometricFocusingDelays}
\end{align}

The excitation time series for each element was a 40-cycle sine pulse with \SI{100}{\kilo\pascal} amplitude at the desired frequency, modulated with a Tukey window with $R = 0.1$, and shifted by the amount given by \cref{eqn:FocusingDelays} [or \cref{eqn:GeometricFocusingDelays} in the uncorrected case].
The relative amplitude for each element was scaled according to the normalized amplitude of $\tilde{p}(x, d)$ (or unity in the geometric case); the maximum amplitude of the corrected case was set to be \SI{20}{\percent} larger than the uncorrected case due to the tapering of the outer elements.
The field was then simulated with each element of the array transmitting with the calculated phase delay and amplitude.

\subsubsection{Passive Acoustic Mapping}
\begin{figure}[!htb]
    \centering
    \includegraphics[width=0.45\textwidth]{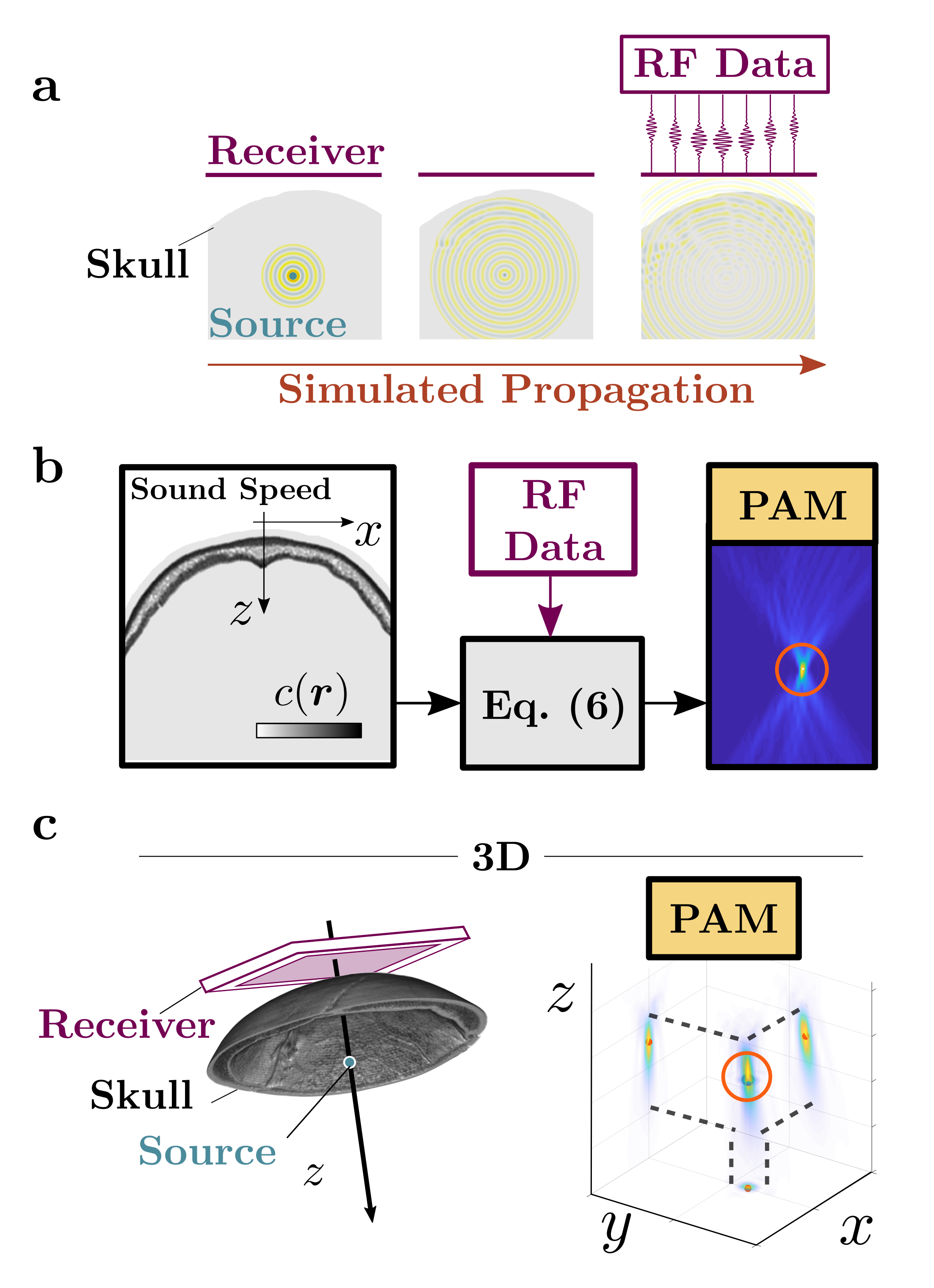}
    \caption{%
      Flow of computations for source localization.
      \textbf{(a)} Trans-skull propagation from the source is simulated, and the the RF data are collected by a virtual array.
      \textbf{(b)}~The known sound speed field $c(\boldsymbol{r})$ is then used with \cref{eqn:AngularSpectrumMarchingAlgorithm} to reconstruct the corrected passive acoustic maps~(PAMs).
      The peak intensity of the PAMs (orange circle) is taken to be the source position, which is compared with the known true location from the simulation. 
      \textbf{(c)}~The reconstructions may be performed in 3D if a planar virtual array is used.
    }
    \label{fig:PAMImplementation}
\end{figure}
Next, the effect of focal aberration correction for the passive acoustic mapping of microbubbles was assessed.
Bubbles were approximated as Gaussian sources with center frequencies of 0.4, 0.8, and 1.2~MHz (to represent harmonic components of bubbles excited with trans-skull FUS at 200~kHz or 600~kHz) \cite{arvanitis_passive_2017,elias_randomized_2016,idbaih_safety_2019}.
Sources ($N = 150$) were placed randomly within a region (approximately 2~mm wide by 3~cm long) representing a vessel shape, within a sound speed environment defined from the CT data (see \cref{fig:PAMAccuracyResults_Frequency}).
To determine the error as a function of position and array aperture, \SI{1}{MHz} sources were then individually simulated in a rectangular grid at axial positions of 30 to 80~mm, and transverse locations from $-$20 to 20~mm.

The resulting pressure from these sources was measured by a virtual linear array with \SI{200}{\micro\meter} pitch (e.g., 250 elements for 50~mm array).
The RF data $\tilde{p}(x, 0)$ were then transformed to give $P_{0}$, from which $P(k_{x}, z)$ was computed with \cref{eqn:AngularSpectrumTransferFunction} or \cref{eqn:AngularSpectrumMarchingAlgorithm} as indicated, with a step size $\Delta z = \SI{50}{\micro\meter}$.
The pressure distribution was then found from $\tilde{p} = \mathcal{F}_{k}^{-1}[P(k_{x}, z)]$, and the pixel intensity field (i.e., the PAM) was computed as $I(x, z) = \| \tilde{p}(x, z) \|^{2}$.
The convolutions in \cref{eqn:AngularSpectrumMarchingAlgorithm} were computed in the frequency domain to improve efficiency, since $P*\Lambda = \mathcal{F}_{k}^{-1}[\tilde{p}\cdot\lambda]$.
Finally, the recovered source positions were taken to be the maxima of the generated PAMs and compared with the known source positions from the simulation.
An algorithm for the reconstruction process is given in the Supplementary Material.

Because $P$ is proportional to $\operatorname{exp}{ik_{z}z}$,  back propagation incurs multiplication by $\operatorname{exp}{-ik_{z}z}$.
Evanescent components of the angular spectrum, for which $k_{z}$ is pure imaginary, will then grow exponentially.
Therefore, all measured angular spectra $P_{0}$ were windowed with a Tukey window with cosine fraction $R = 0.25$ to taper these components \cite{williams_fourier_1999}. 
Additionally, all initial spectra were zero padded such that their computational extent was four times larger than their physical extent.
Sound speed fields $c(\boldsymbol{r})$ were padded with their edge values replicated to match the grid size of the padded $P_{0}$.
The axial step $\Delta z$ was \SI{50}{\micro\meter} and the field was computed from $z = 0$ to \SI{90}{\milli\meter}.


\section{Results}
\label{sec:Results}

\subsection{Focal Aberration Correction}
\label{sec:FocalAberrationCorrection}
\begin{figure*}[!htb]
    \centering
    \includegraphics[width=0.75\textwidth]{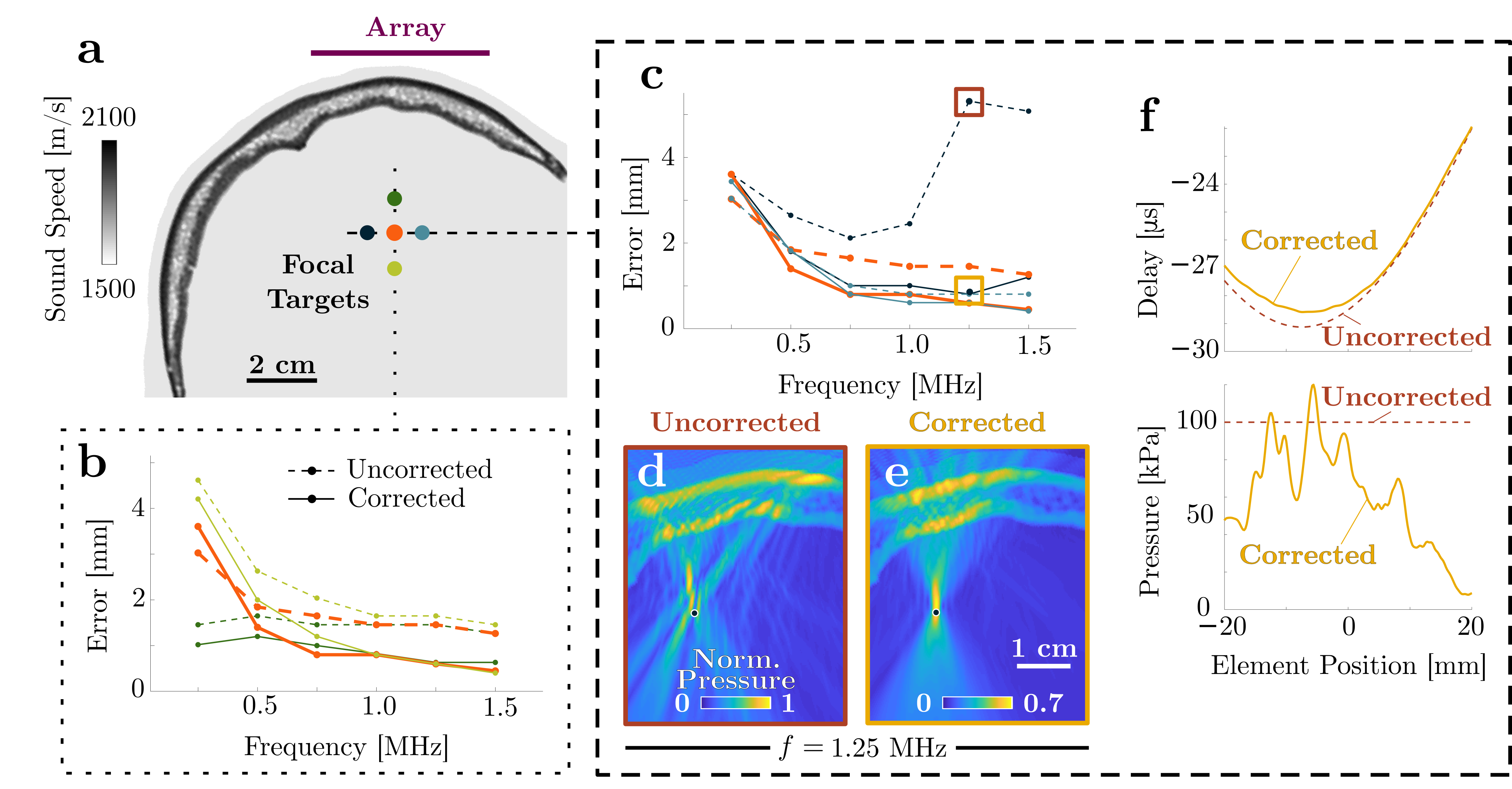}
    \caption{Use of phase corrections and amplitude shading enable improved accuracy in focal targeting. %
    \textbf{(a)}~Geometry for the simulation. %
    Error in the position of the maximum simulated pressure compared with the target focus point with (solid lines) and without (dashed lines) the phase and amplitude corrections at different 
    \textbf{(b)}~axial and 
    \textbf{(c)}~transverse focal targets. 
    For the outlying case at \SI{1.25}{MHz} in \textbf{(c)}, the 
    \textbf{(d)}~uncorrected and
    \textbf{(e)}~corrected maximum pressure distributions (normalized to their respective maxima) are shown, computed with
    \textbf{(f)}~the phase corrected and uncorrected time delays and amplitude shadings.
    }
    \label{fig:FocusingErrorResults}
\end{figure*}

First we estimated the effect of the phase correction on transmit focal accuracy for various focal targets as a function of frequency (\cref{fig:FocusingErrorResults}).
For each target focus, the focal delays were computed geometrically with \cref{eqn:GeometricFocusingDelays} and then with corrections given by \cref{eqn:AngularSpectrumMarchingAlgorithm,eqn:FocusingDelays}. 
Across all focal positions and frequencies, the focal error was reduced from \SI{2.1 \pm 1.2}{\milli\meter} without phase corrections to \SI{1.3 \pm 1.0}{\milli\meter} with the correction.
Aberration errors in focal targeting were generally larger at off-axis focal positions. 
See, for example the highlighted case in \cref{fig:FocusingErrorResults}(c--e), where the uncorrected focal error was \SI{5.1}{\milli\meter}, while the error for the corrected case was \SI{0.6}{\milli\meter}.

We also evaluated the impact of the standoff distance of the array form the skull on focusing. 
Results indicate that the focal accuracy and improvement were comparable for different standoff distances (\SI{0.6\pm0.3}{\milli\meter} with the correction and \SI{1.6\pm1.3}{\milli\meter} without, see Supplementary Fig.~S-1).

\begin{figure}[!htb]
    \centering
    \includegraphics[width=0.46\textwidth]{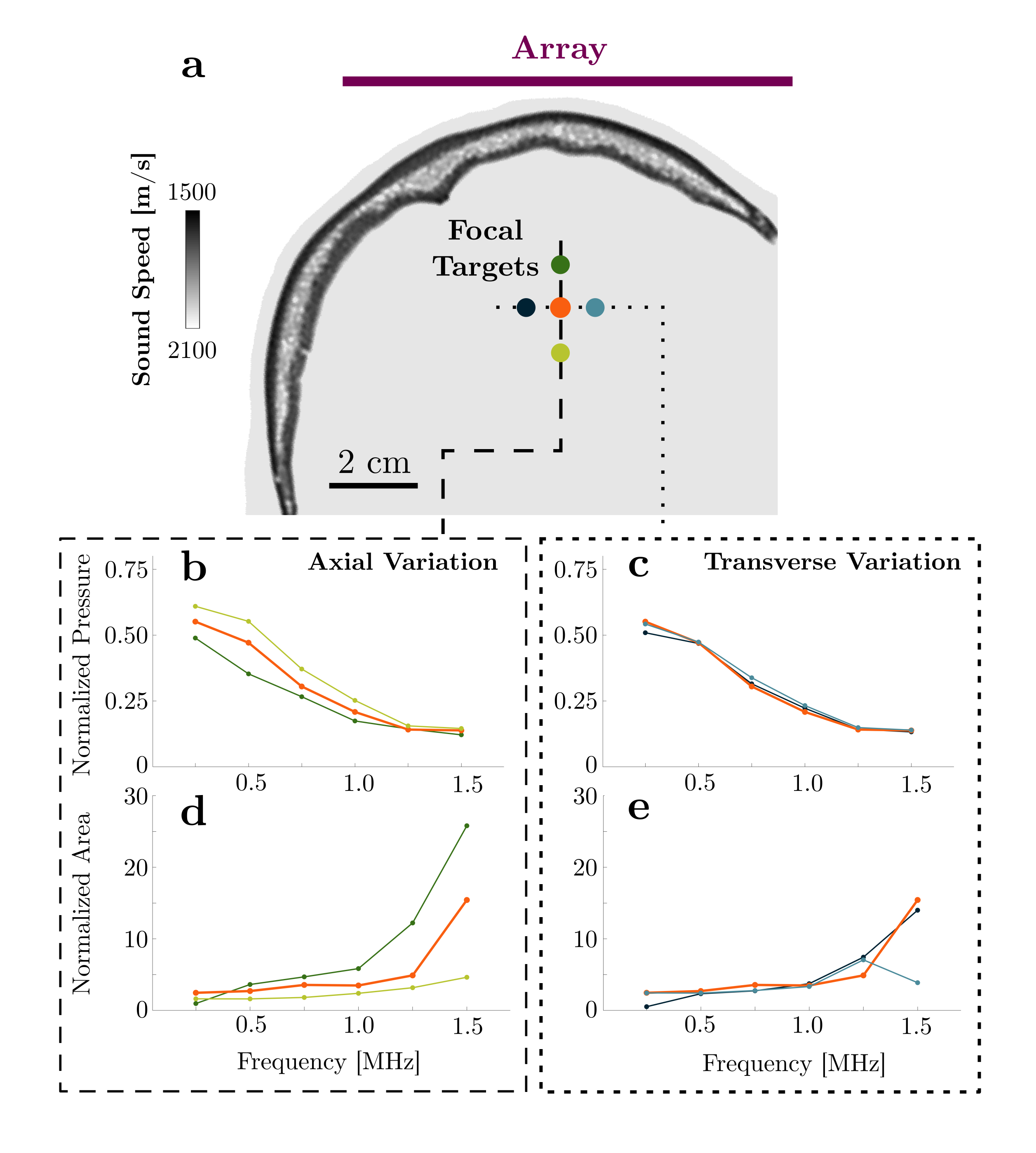}
    \caption{Corrected transmit focusing achieves focal pressures and focal spot sizes comparable to the free field case. %
    \textbf{(a)}~Geometry for the simulation. %
    \textbf{(b)}~Axial and
    \textbf{(c)}~transverse variation in peak pressure amplitude as a function of frequency, normalized to amplitude obtained in simulations in water.
    \textbf{(d)}~Axial and
    \textbf{(e)}~transverse focal area (\SI{3}{dB}) normalized to the spot size obtained in the simulated water case.
    }
    \label{fig:FocusingParameters}
\end{figure}

Next we assessed the normalized (to free field) focal pressure and spot size (i.e., the total area within \SI{3}{dB} of the peak pressure) achieved with the proposed aberration correction algorithm as a function frequency and position.
\Cref{fig:FocusingParameters} illustrates how the spot size and focal pressure vary in the corrected case compared with the water path case as a function of frequency and position.
The decrease in maximum pressure with frequency that is observed [\cref{fig:FocusingParameters}(b--c)] reflects absorption losses that  increase as $f^{\beta}$ ($\beta = 1.2$ was used).
Relative to the respective water case, the area of the focal region in the corrected trans-skull case did increase higher frequencies; however the absolute area became smaller, from \SI{22.9\pm13.8}{mm\squared} for \SI{250}{kHz} to \SI{6.5\pm3.7}{mm\squared} for the \SI{1.5}{MHz} case.
Both the spot size and focal pressure were seen to  vary more with the focus' axial position than with its transverse position [compare \cref{fig:FocusingParameters}(b\&d) and \cref{fig:FocusingParameters}(c\&e)]. 
The focal error and spot size in the corrected and uncorrected cases averaged over all frequencies and positions are summarized in  \cref{tab:FocalErrorSkull}.
Together, these data demonstrate that the proposed method is able to correct aberrations and improve focal targeting through the human skull.

\begin{table}[!htb]
  \renewcommand{\arraystretch}{1.2}
    \caption{Mean and standard deviation focal location error and spot size with uncorrected [\cref{eqn:FocusingDelays}] and the corrected [\cref{eqn:GeometricFocusingDelays}] focusing delays, averaged over all frequencies and positions shown in \cref{fig:FocusingErrorResults,fig:FocusingParameters}.}
  \label{tab:FocalErrorSkull}
  \centering
  \begin{tabular}{ccccc}
           \,               & \multicolumn{2}{c}{\textbf{Uncorrected}} & \multicolumn{2}{c}{\textbf{Corrected}} \\
    \textbf{Aperture} & \textbf{Error [mm]} & \textbf{Size [mm\textsuperscript{2}]} & \textbf{Error [mm]} & \textbf{Size [mm\textsuperscript{2}]} \\
    \hline
    \textbf{50~mm}             & 2.1 $\pm$ 1.2       &  6.8 $\pm$ 5.7      & 1.3  $\pm$ 1.0  & 12.6 $\pm$ 16.9 \\
    \textbf{100~mm}            & 2.2 $\pm$ 0.7       &  7.5 $\pm$ 6.4     & 1.0  $\pm$ 0.4  & 9.6 $\pm$ 10.9             
  \end{tabular}
\end{table}

\subsection{Passive Acoustic Mapping}
\label{sec:PassiveAcousticMapping}
\begin{figure*}[!htb]
    \centering
    \includegraphics[width=0.95\textwidth]{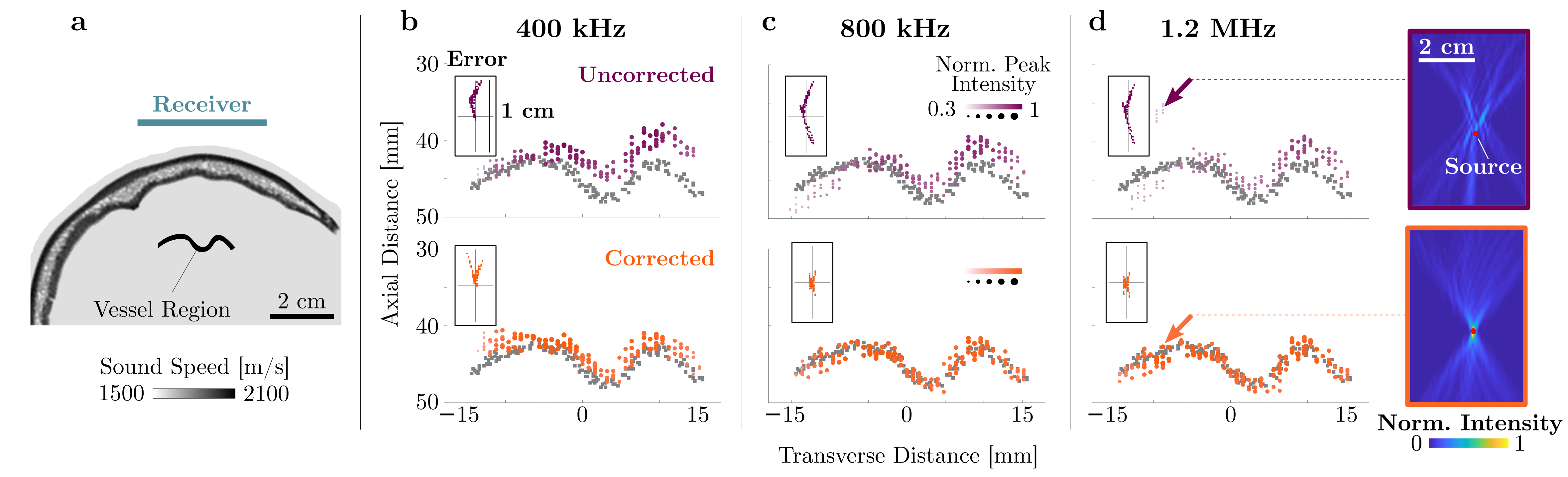}
    \caption{%
     Accuracy of source localization for uncorrected and phase corrected trans-skull ASA PAM reconstructions.
     \textbf{(a)}~Geometry of the receiver and region in which sources were randomly placed.
     In \textbf{(b--d)}, gray squares represent true source positions, and circles indicate recovered location. 
     The size and opacity of each circle indicates the intensity of the peak in the PAM, normalized to the maximum intensity for all peaks at that frequency.
    The insets plot each recovered source position relative to its truth position, and represent the distribution of axial and transverse errors over all reconstructions for that frequency.
     Peaks from PAMs computed with unmodified algorithm \cref{eqn:AngularSpectrumTransferFunction} (purple) and corrected
      algorithm \cref{eqn:AngularSpectrumMarchingAlgorithm} (orange) for 
      \textbf{(b)}~$f = \SI{400}{\kilo\hertz}$, 
      \textbf{(c)}~$f = \SI{800}{\kilo\hertz}$, and 
      \textbf{(d)}~$f = \SI{1.2}{\mega\hertz}$. 
      In \textbf{(d)}, the normalized PAMs for the sources indicated by the arrows are shown for comparison with the true source position (red dot).
      Aperture for all simulations was 50~mm.%
    }
    \label{fig:PAMAccuracyResults_Frequency}
\end{figure*}

To assess the ability of the proposed method for PAM we estimated the errors in the peak location compared with the source position for three different frequencies (\Cref{fig:PAMAccuracyResults_Frequency}).
The total (radial) error using the heterogeneous ASA was reduced compared to that for the PAMs based on the homogeneous medium (i.e., uncorrected) ASA. 
Specifically, using \cref{eqn:AngularSpectrumMarchingAlgorithm}, the localization errors were \SI{1.6 \pm 0.9}{\milli\meter}, \SI{0.6 \pm 0.4}{\milli\meter}, and \SI{0.6 \pm 0.5}{\milli\meter} for \SI{400}{\kilo\hertz}, \SI{800}{\kilo\hertz}, and \SI{1.2}{\mega\hertz} sources, respectively, compared with \SI{2.8 \pm 1.5}{\milli\meter}, \SI{2.0 \pm 1.1}{\milli\meter}, and \SI{2.2 \pm 1.8}{\milli\meter} when \cref{eqn:AngularSpectrumTransferFunction} was used (i.e., when the presence of the skull was not accounted for).
Importantly, across all simulated point sources, there were  no outliers in the corrected case [e.g., in \Cref{fig:PAMAccuracyResults_Frequency}(d), where the PAMs for the indicated sources are shown at right].
Further, the intensities of the peaks in the reconstructions were larger and more uniform in the corrected case overall: normalized to the mean corrected peak intensity for each case, the uncorrected PAMs had peak intensities of 0.94$\,\pm\,$0.09, 0.78$\,\pm\,$0.12, and 0.59$\,\pm\,$0.14 for the \SI{400}{\kilo\hertz}, \SI{800}{\kilo\hertz}, and \SI{1.2}{\mega\hertz}, cases respectively.

\begin{figure*}[!htb]
    \centering
    \includegraphics[width=0.72\textwidth]{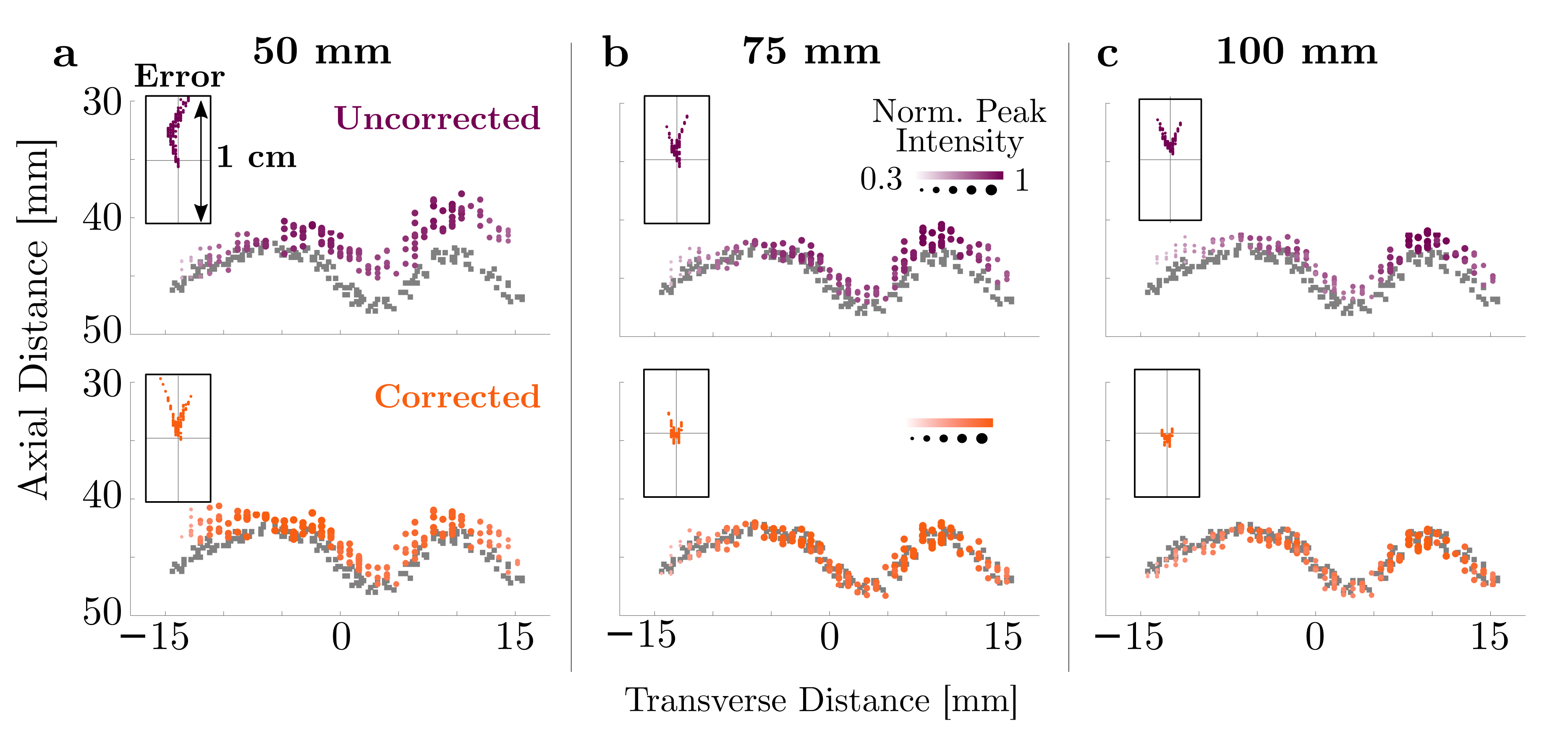}
    \caption{
     Phase corrected trans-skull ASA PAM reconstructions yield improved source localization accuracy and intensity compared to the uncorrected case.
     In \textbf{(a--c)}, gray squares represent true source positions, and circles indicate recovered location. 
     The size and opacity of each circle indicates the intensity of the peak in the PAM, normalized to the maximum intensity for all peaks at that frequency.
     The insets plot each recovered source position relative to its truth position, and represent the distribution of axial and transverse errors over all reconstructions for that frequency.
     Peaks from PAMs computed with unmodified algorithm \cref{eqn:AngularSpectrumTransferFunction} (purple) and corrected
      algorithm \cref{eqn:AngularSpectrumMarchingAlgorithm} (orange) for apertures of
      \textbf{(a)}~\SI{50}{\milli\meter}, 
      \textbf{(b)}~\SI{75}{\milli\meter}, and 
      \textbf{(c)}~\SI{100}{\milli\meter}. 
      The frequency of all simulations sources was 400~kHz.}
    \label{fig:PAMAccuracyResults_Aperture}
\end{figure*}

Next we assessed the localization accuracy as a function of the aperture size. 
For larger apertures with the same pitch (i.e., with additional elements), the localization is further improved for the corrected case. 
\Cref{fig:PAMAccuracyResults_Aperture} shows the effect of the aperture size on localization accuracy for the trans-skull case with the uncorrected ASA (top row) and corrected ASA (bottom row) for sources at 400~kHz.
For the corrected case, the localization error decreases with increasing aperture size (\SI{1.5 \pm 0.9}{\milli\meter}, \SI{0.4 \pm 0.3}{\milli\meter}, and \SI{0.5 \pm 0.2}{\milli\meter} for \SI{50}{\milli\meter}, \SI{75}{\milli\meter}, and \SI{100}{\milli\meter} apertures, respectively). 
For the uncorrected case, a larger aperture does not reduce the error [\SI{2.7 \pm 1.6}{\milli\meter}, \SI{1.2 \pm 0.8}{\milli\meter}, and \SI{1.5 \pm 0.6}{\milli\meter} for \SI{50}{\milli\meter}, \SI{75}{\milli\meter}, and \SI{100}{\milli\meter} apertures, respectively; see \cref{fig:PAMAccuracyResults_Aperture}(b--c)].

\begin{figure*}[!htb]
    \centering
    \includegraphics[width=0.75\textwidth]{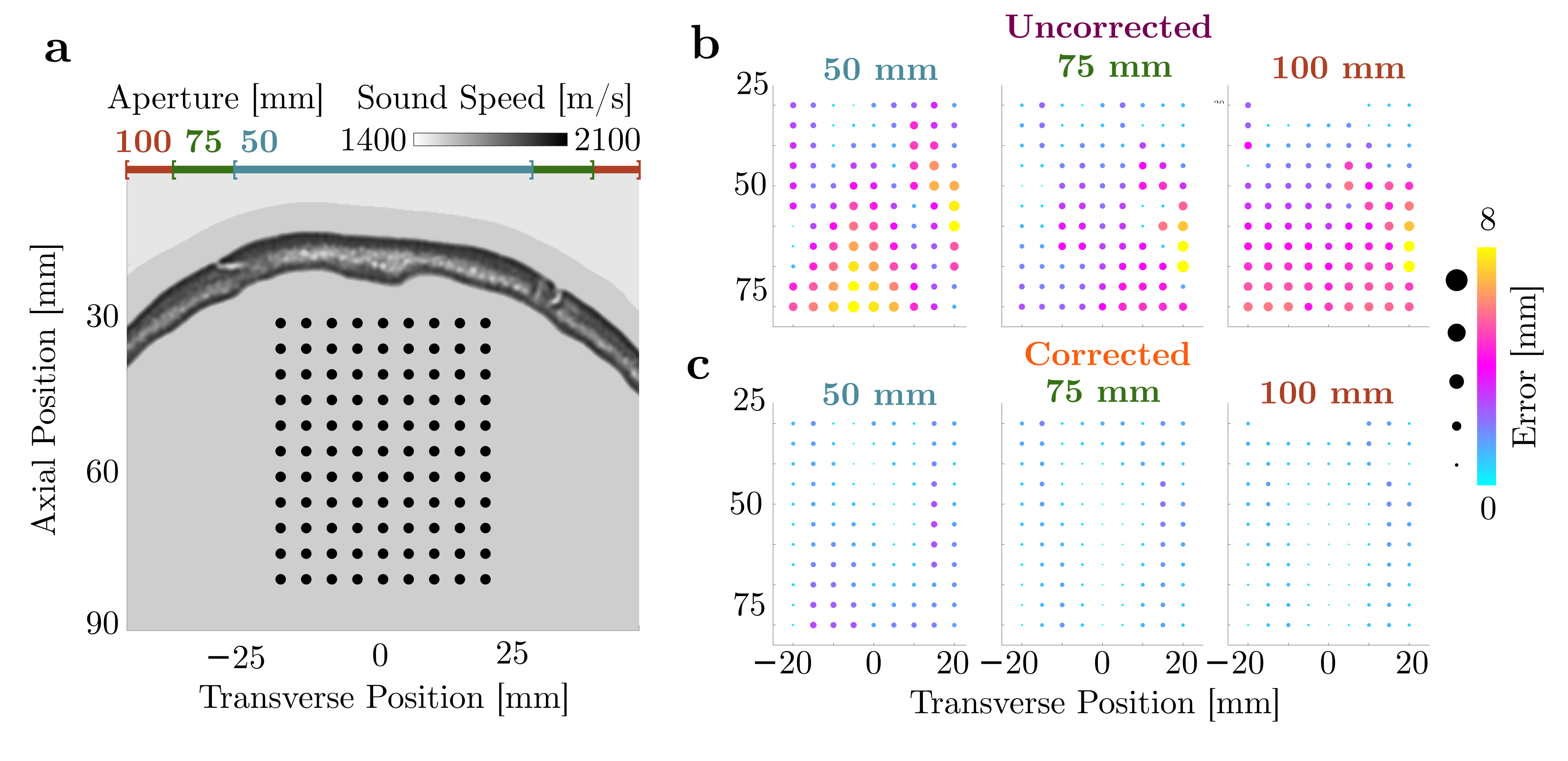}
    \caption{%
      Source localization accuracy for phase corrected trans-skull ASA PAM reconstructions improves over 40~mm by 50~mm field of view compared to the uncorrected case for \SI{1}{\mega\hertz} sources.
      \textbf{(a)}~Sound speed field computed from CT data for a human skull, and relative position of the simulated sources (black circles) and array.
      Error between the location computed from the maximum value of the PAM  and the true source location, for the indicated aperture size,
      \textbf{(b)}~formed with \cref{eqn:AngularSpectrumTransferFunction} and
      \textbf{(c)}~with \cref{eqn:AngularSpectrumMarchingAlgorithm}.
    }
    \label{fig:PAMAccuracyResultsSkull}
\end{figure*}

Finally, we estimated the localization error as a function of position and aperture using a rectangular grid of \SI{1}{MHz} sources within the skull (\Cref{fig:PAMAccuracyResultsSkull})
Across all positions, the localization error was reduced by \SIrange{60}{80}{\percent}, and the localization error in the corrected positions decreased monotonically with aperture for the phase corrected case.
Further, the peak intensity in the uncorrected case was on average \SIrange{30}{40}{\percent} smaller than in the corrected case.
\Cref{tab:LocalizationErrorSkull} summarizes the mean and standard deviation of the error over all points in the grid in \cref{fig:PAMAccuracyResultsSkull}.
Full details, including the axial and transverse error and spot size are included in the Supplementary Tables~S-1--S-3.

\begin{table}[!htb]
  \renewcommand{\arraystretch}{1.2}
    \caption{Mean and standard deviation PAM localizaion error and peak intensities with uncorrected [\cref{eqn:AngularSpectrumTransferFunction}] and the corrected [\cref{eqn:AngularSpectrumMarchingAlgorithm}] beamforming, computed over the positions shown in \cref{fig:PAMAccuracyResultsSkull}.}
  \label{tab:LocalizationErrorSkull}
  \centering
  \begin{tabular}{ccccc}
           \,               & \multicolumn{2}{c}{\textbf{Uncorrected}} & \multicolumn{2}{c}{\textbf{Corrected}} \\
    \textbf{Aperture} & \textbf{Error [mm]} & \textbf{Intensity} & \textbf{Error [mm]} & \textbf{Intensity} \\
    \hline
    \textbf{50~mm}             & 3.7 $\pm$ 2.2       &  0.73 $\pm$ 0.46     & 1.2  $\pm$ 0.7  & 1.0 $\pm$ 0.60 \\
    \textbf{75~mm}             & 2.5 $\pm$ 1.7       &  0.66 $\pm$ 0.36     & 0.9  $\pm$ 0.5  & 1.0 $\pm$ 0.47 \\
    \textbf{100~mm}            & 3.5 $\pm$ 1.9       &  0.62 $\pm$ 0.28     & 0.8  $\pm$ 0.4  & 1.0 $\pm$ 0.39             
  \end{tabular}
\end{table}

\subsection{Computational Efficiency}
\label{sec:Efficiency}
For focal aberration correction, calculation of the focusing delays and amplitudes from \cref{eqn:AngularSpectrumMarchingAlgorithm,eqn:FocusingDelays} required \SI{25.2 \pm 8.5}{\milli\second}. 
Note that registration of the the 2D sound speed map with the computational grid required \SI{1.63 \pm 0.03}{\second}, but only needs to be done once during registration.
For PAM, the correction using the heterogeneous ASA ([\cref{eqn:AngularSpectrumMarchingAlgorithm}]) the reconstructions took approximately \SI{166 \pm 37}{\milli\second}, for a step size $\Delta z = \SI{50}{\micro\meter}$ over all calculations. 
This is approximately four times slower than the uncorrected ASA [i.e., with \cref{eqn:AngularSpectrumTransferFunction}], where the computation of the PAMs required approximately \SI{44 \pm 4}{\milli\second} per frequency.
For a 3D PAM example case, \SI{540 \pm 18}{\milli\second} was required per frequency to evaluate \cref{eqn:AngularSpectrumMarchingAlgorithm} for data from a 24-by-24 element array and \SI{100}{\micro\meter} step size (173\,376 voxels) and 3054 time samples.
All reported times are for a standard desktop computer with no parallel processing or GPU acceleration.


\section{Discussion}
\label{sec:Discussion}
This work augments the ASA for sound propagation in heterogeneous media with a fast phase correction technique (\cref{fig:FocusingImplementation,fig:PAMImplementation}).
The methods were derived under assumptions of relatively weak heterogeneity, specifically that $\left| \nabla\rho_{0}/\rho_{0} + 2\nabla c/c \right|$ is small compared with the wavelength (see Supplementary Material).
While these conditions are not strictly met by the impedance contrast represented by the skull, our results from simulations with clinical CT data have demonstrated more than \SI{50}{\percent} reduction in localization and transmission focusing error at clinically relevant frequencies and geometries.
This improvement in accuracy is on the scale of a half wavelength with little additional computational burden. 

Via phase extraction and amplitude shading, the error in focal targeting was reduced by approximately \SI{60}{\percent} as compared to the uncorrected case (from \SI{2.2 \pm 0.7}{\milli\meter} without phase corrections to \SI{1.0 \pm 0.4}{\milli\meter} with the full correction---over a range of positions and frequencies, \cref{fig:FocusingErrorResults}). 
The improvement in accuracy  is more pronounced at higher frequencies, since the derivation assumes $c(\boldsymbol{r})$ varies slowly compared with the wavelength; thus the approximation is more valid for smaller wavelengths (higher frequencies). 
This is not considered as a major impediment, as aberration is more pronounced at higher frequencies, where the performance of the proposed method is favorable (\cref{fig:FocusingParameters}).

Further, the method was seen to reduce the error in trans-skull source localization from computed PAMs by \SIrange{60}{80}{\percent}, (from \SI{2.5 \pm 2.3}{\milli\meter} to \SI{0.6\pm0.5}{\milli\meter}) for a \SI{1.2}{MHz} sources (\cref{fig:PAMAccuracyResults_Frequency}). 
Localization accuracy was seen to increase with larger apertures (\cref{fig:PAMAccuracyResults_Aperture}). 
A larger aperture imparts higher spatial frequency resolution (i.e., smaller $\Delta k_{x}$ and $\Delta k_{y}$ in the discrete transform), and thus improved coherence of the reconstruction at the correct source location.
Larger apertures in the uncorrected case do not in general improve the accuracy of the localization in the presence of aberrations. 

Compared with previously proposed methods for adaptation of ASA to heterogeneous media, our method has the advantage of requiring a single propagation step, and does not require transformations between each axial position \cite{vyas_ultrasound_2012,clement_non-invasive_2002,clement_forward_2003}.
Its inherent efficiency should allow its use in conjunction with real-time control methods based on PAM that allow for location-specific cavitation control \cite{patel_closed_2018}. 
Moreover, the proposed algorithm is faster than full simulation methods for phase correction (e.g., 166 ms, compared with 2 minutes for the corresponding 2D k-Wave simulation).
Three dimensional heterogeneous PAM examples required 520~ms per frequency for $1.7\times10^{5}$ voxels with 576 channels and 3054 time steps, which is comparable to GPU-optimized homogeneous time domain algorithms run on high performance computers (e.g., 85~ms for $10^{3}$~voxels and $3\times10^{5}$ time samples in Ref.~\citenum{jones_three-dimensional_2018}).

Computationally efficient methods may also lead to improved aberration correction accuracy by allowing iterative focal improvement. 
Such approaches are further supported by clinical experience that has shown that one still needs to correct for small errors (\SIrange{1}{2}{mm}) during targeting. \cite{elias_pilot_2013,elias_randomized_2016}
Currently, these corrections are calculated from CT data and confirmed via MRI–guided methods, wherein low-level focal heating (a few degrees Celsius) is measured with MR temperature imaging~(MRTI) \cite{martin_high-intensity_2009}.
In experimental settings, aberration correction has also been confirmed by small tissue displacements (a few microns) that are induced by radiation force, using MR acoustic radiation force imaging~(MR-ARFI) \cite{kaye_adapting_2013,mcdannold_magnetic_2008}. 
Based on these methods, optimization routines can be developed to refine the phase of each element by maximizing focal displacement \cite{marsac_mr-guided_2012,vyas_transcranial_2014} or focal heating for a given excitation pulse.
To ensure that such adaptive focusing methods are performed in timescales similar to current CT-based methods for a single correction (of the order of seconds) the computation of the phase delays needs to take place on time scales of the order of a few milliseconds.
Hence the proposed methods can also provide fast PAM for spatio-temporal control of the cerebrovascular microbubble dynamics.

Our study has some limitations.
First, only a single skull has been used on our simulations; in future work we aim to evaluate the proposed approach using different skulls and skull segments. 
Additionally, experimental validation using different skulls, microbubbles, and phased arrays is needed for assessing the robustness of the proposed method for focal aberration correction and imaging (passive or active). 
Further, the algorithm is derived under the assumption of forward propagation only;
extension of the method to account for reflection is nontrivial, but could improve the localization and focal targeting accuracy.
The reported computation times have strong dependence on the hardware, size of the domain, and the choice of reconstruction parameters (e.g., step size, number of pad channels, etc.). 
However, in all cases presented it could be reduced to a few milliseconds with more sophisticated implementations, including more powerful hardware, parallel computing, and graphical processing techniques.
Finally, our formulation does not include reflection, attenuation, and nonlinearity; 
however, the proposed method can be readily augmented to account for the effects of  attenuation,\cite{gu_numerical_2018} and nonlinearity\cite{jing_evaluation_2011,jing_improved_2014}  potentially enabling the quantification of the acoustic emissions and the investigation of nonlinear sound propagation though heterogeneous media, albeit at the price of some additional computational burden.


\section{Conclusion}
\label{sec:Conclusion}
We derived the general solution for the heterogeneous ASA. 
Numerical data showed that the general solution provides accurate trans-skull transmit focusing and point source localization.  
Sub-millimeter errors were attained with only a modest increase in computational complexity both for focusing and point source localization using clinically relevant frequencies (\SIrange{0.25}{1.5}{\mega\hertz}) and array apertures (\SIrange{50}{100}{\milli\meter}). 
The computation times on a standard computer were comparable to those reported for GPU-optimized uncorrected time domain algorithms on high performance computers.
Collectively, our findings indicate that the heterogeneous ASA may create new possibilities for treatment, treatment monitoring, and diagnosis of brain diseases.

\begin{appendices}
\numberwithin{equation}{section}
\numberwithin{figure}{section}
\setcounter{equation}{0}
\setcounter{figure}{0}
{ \small

\section{Derivation of ASA Results}
\label{sec:Derivations}
\subsection{Governing Equation}
Provided the medium is weakly heterogeneous (see Supplementary Material), propagation may be described by
\begin{align}
  \nabla^{2}p - \frac{1}{c^{2}(\boldsymbol{r})}\ppd{p}{t} &= 0\,.
  \label{eqn:InhomogeneousWaveEquation}
\end{align}
The sound speed may be written as the sum of a reference sound speed $c_{0}$ and a spatially-varying part $c'(\boldsymbol{r})$\cite{agrawal_angular_1975}
\begin{align}
  c(\boldsymbol{r}) &= c_{0} + c'(\boldsymbol{r}).
  \label{eqn:SoundSpeedVariance}
\end{align}
Because \cref{eqn:InhomogeneousWaveEquation} is most valid for small changes in sound speed \cite{bergmann_wave_1946}, the mean value of $c(\boldsymbol{r})$ is the natural choice for the reference sound speed $c_{0}$.
Defining $\mu(\boldsymbol{r}) = c_{0}^{2}/c^{2}(\boldsymbol{r})$
and taking the temporal Fourier transform of \cref{eqn:InhomogeneousWaveEquation} gives 
\begin{align}
 \left( \nabla^{2} + k_{0}^{2} \right)\tilde{p} &=  k_{0}^{2}\left(1 - \mu \right)\tilde{p}\,,
  \label{eqn:InhomogeneousMediumHelmholtz}
\end{align}
where $k_{0} = \omega/c_{0}$. 
Note that for a uniform medium, then $\mu = 1$, and Eq.~(\ref{eqn:InhomogeneousMediumHelmholtz}) reduces to the homogeneous Helmholtz equation as expected. 
Now defining an auxiliary function $\lambda(\boldsymbol{r}) \equiv k_{0}^{2}\left(1 - \mu \right)$, \cref{eqn:InhomogeneousMediumHelmholtz} may be written
\begin{align}
  \left( \nabla^{2} + k_{0}^{2} \right)\tilde{p}
  &=
  \mathcal{F}^{-1}_{k}\left[ \Lambda * P \right]\,,
  \label{eqn:AngularSpectrumResult_Transforms}
\end{align}
where the convolution theorem has been used, and $\Lambda(k_{x}, k_{y}, z) = \mathcal{F}_{k}[\lambda(x, y, z)]$.
Then, since
\begin{align}
   \mathcal{F}_{k}\left[\nabla^{2} \tilde{p}\,\right] 
   = 
   \left(-k_{x}^{2} - k_{y}^{2} + \ppd{}{z}\right)P\,,
   \label{eqn:SpatialDifferentiationTheorem}
\end{align}
the left hand side of \cref{eqn:InhomogeneousMediumHelmholtz} becomes
\begin{align}
  \left( \nabla^{2} + k_{0}^{2} \right)\tilde{p} 
  &= 
  \nabla^{2}\tilde{p} + k_{0}^{2}\tilde{p} 
  \nonumber \\
  &= \mathcal{F}_{k}^{-1}\,\left[\left( - k_{x}^{2} - k_{y}^{2} + \ppd{}{z} \right)P + k_{0}^{2}P \right] 
  \nonumber \\
  &= \mathcal{F}_{k}^{-1}\,\Bigg[ \Bigg( \underbrace{k_{0}^{2} - k_{x}^{2} - k_{y}^{2}}_{=\,k_{z}^{2}} + \ppd{}{z} \Bigg)P \Bigg]\,.
  \label{eqn:GoverningEquationLHS}
\end{align}
Then \cref{eqn:AngularSpectrumResult_Transforms,eqn:GoverningEquationLHS} yield \Cref{eqn:InhomogeneousGoverningEquation}
\begin{align}
  \mathcal{F}^{-1}_{k}\left[ \ppd{P}{z} + k_{z}^{2}P\right] 
  &= 
  \mathcal{F}^{-1}_{k}\left[ \Lambda * P \right]
  \nonumber \\
  &\hspace{-3mm}\implies 
  \ppd{P}{z} + k_{z}^{2}P 
  = 
  \Lambda * P\,.
  \label{eqn:InhomogeneousGoverningEquation_apx}
\end{align}

\subsection{Implicit Solution and Approximation}
\Cref{eqn:InhomogeneousGoverningEquation_apx} may be approached in a general way by treating the right hand side as a source term and using a Green's function technique \cite{jing_evaluation_2011}.
The homogeneous solution is
\begin{align}
  P_{h} = Ae^{ik_{z}z} + Be^{-ik_{z}z}\,,
  \label{eqn:AngularSpectrumOdeHomogeneousSolution}
\end{align}
and the appropriate Green's function for the 1D homogeneous Helmholtz equation is\cite{watanabe_integral_2015,morse_methods_1953}
\begin{align}
  g(z\,|\,z') 
  &=
  \frac{1}{2ik_{z}}\left( e^{ik_{z}| z - z' |} - e^{ik_{z}| z + z' |}\right)\,.
\end{align}
Then the full solution is then given by
\begin{align}
  P 
  &=
  P_{h}
  +
  \int_{0}^{\infty}{ g(z\,|\,z')\times \Lambda * P\,\mathrm{d}z' }\,.
\end{align}
If it is assumed that there are no backward travelling waves, then $B = 0$, and from the boundary condition $P|_{z = 0} = P_{0}$ at the source plane, the implicit solution is found [\cref{eqn:AngularSpectrumSolutionImplicit}, repeated here]:
\begin{align}
  P 
  &=
  P_{0}e^{ik_{z}z}
  +
  \frac{\,e^{ik_{z}z}}{2ik_{z}}\int_{0}^{z}{ e^{-ik_{z}z'}\left( \Lambda * P  \right)\,\mathrm{d}z' }\,.
  \label{eqn:AngularSpectrumSolutionImplicit_apx}
\end{align}
The approximate marching scheme suggested by Jing et al. \cite{jing_improved_2014} defines the angular spectrum at some discrete axial position $z_{n}$ to be
\begin{align}
    P^{n} = P(k_{x}, k_{y}, z_{n})\,,
\end{align}
and
\begin{align}
  P^{n + 1} = P(k_{x}, k_{y}, z_{n} + \Delta z).
\end{align}
Then, with knowledge of the initial condition $P^{0} = P_{0}$, the field may be approximated at arbitrary $z$ via
\begin{align}
  P^{n + 1} 
  \approx 
  P^{n}e^{ik_{z}\Delta z}
  +
  \frac{e^{ik_{z}\Delta z}}{2ik_{z}}\,\left( P^{n} * \Lambda  \right) \times \Delta z\,.
  \label{eqn:AngularSpectrumMarchingAlgorithm_apx}
\end{align}

}
\end{appendices}

\flushleft
\nomenclature[A, f]{$f$}{Frequency}
\nomenclature[A, p]{$p$}{Acoustic pressure}
\nomenclature[A, c]{$c$}{Sound speed}
\nomenclature[A, c]{$c_{0}$}{Reference (average) sound speed}
\nomenclature[A, k]{$k_{i}$}{Wavenumber component}
\nomenclature[A, r]{$\boldsymbol{r}$}{Position vector $\boldsymbol{r} = (x, y, z)$}
\nomenclature[A, g]{$g$}{Green's function}
\nomenclature[A, p]{$P$}{Angular spectrum (spatial Fourier transform) of harmonic pressure ($\mathcal{F}_{k}[\tilde{p}] = P$)}
\nomenclature[A, p]{$\tilde{p}$}{Harmonic pressure amplitude ($\mathcal{F}[p] = \tilde{p}$)}
\nomenclature[A, R]{$R$}{Tukey window taper parameter ($R = 1$ for Hann window, $R = 0$ for rectangular window). }
\nomenclature[A, k]{$k_{0}$}{Reference wavenumner $k_{0}^{2} = (\omega/c_{0})^{2} = k_{x}^{2} + k_{y}^{2} + k_{z}^{2}$. }

\nomenclature[G]{$\alpha$}{Attenuation}
\nomenclature[G]{$\omega$}{Angular frequency $\omega = 2\pi f$}
\nomenclature[G]{$\rho$}{Acoustic density}
\nomenclature[G]{$\tau$}{Time delay}
\nomenclature[G]{$\mu$}{Sound speed coefficient $ \mu(\boldsymbol{r}) \equiv c_{0}^{2}/c^{2}(\boldsymbol{r})$}
\nomenclature[G]{$\lambda$}{Auxiliary function $\lambda \equiv k_{0}^{2}(1 - \mu)$}
\nomenclature[G]{$\Lambda$}{Angular spectrum of auxiliary function $\Lambda = \mathcal{F}_{k}[\lambda]$}
\nomenclature[G]{$\beta$}{Absorption power law exponent}

\nomenclature[M, d]{$\nabla$}{Gradient operator}
\nomenclature[M, d]{$\nabla\cdot$}{Divergence operator}
\nomenclature[M, d]{$\nabla^{2}$}{Laplacian operator}
\nomenclature[M, f]{$\mathcal{F}$}{Temporal Fourier transform $\mathcal{F}[\,\cdot\,] \equiv \int_{-\infty}^{\infty}{(\cdot)\,e^{i\omega t}\,\mathrm{d}t}$}
\nomenclature[M, fk]{$\mathcal{F}_{k}$}{Spatial Fourier transform $\mathcal{F}_{k}[\,\cdot\,] \equiv \iint_{-\infty}^{\infty}{(\cdot)\,e^{-i(k_{x}x + k_{y}y)}\,\mathrm{d}x\,\mathrm{d}y}$}
\nomenclature[M, c]{$*$}{2D convolution $f*g \equiv \iint_{-\infty}^{\infty}{f(k_{x} - k_{x}',\,k_{y} - k_{y}',\,z)\,g(k_{x}, k_{y}, z)\,\mathrm{d}k_{x}'\,\mathrm{d}k_{y}'}$}

\printnomenclature[12mm] \vspace{5mm}%

\bibliographystyle{IEEEtran6}
\bibliography{main_arxiv}

\end{document}